\documentclass[showpacs,preprintnumbers,amsmath,amssymb,aps, prd]{revtex4}
\usepackage{graphicx}
\usepackage{grffile}
\usepackage{mathrsfs, mathtools}
\usepackage{caption}
\usepackage{float}
\usepackage{xcolor}
\usepackage{changepage}
\usepackage{dcolumn}
\usepackage{bm}
\usepackage{graphicx,subfigure,epsfig}
\usepackage{multirow}

\def\mn{_{\mu\nu}}
\def\MN{^{\mu\nu}}

\def\a{\alpha}
\def\b{\beta}
\def\f{\frac}

\def\l{\mathcal{L}}

\def\ab{\overline{a}}

\def\bb{\overline{b}}
\def\tb{\tilde{\beta}}
\def\ta{\tilde{\alpha}}
\def\tg{\tilde{g}}
\def\c{\cite}
\def\r{\ref}
\def\ar{A(r)}
\def\br{B(r)}
\def\fx{F(x)}
\def\hx{H(x)}
\def\s{Schwarzschild }
\def\h{Hernquist DM halo }
\newcommand\be{\begin{equation}}
\newcommand\ee{\end{equation}}
\newcommand\ba{\begin{eqnarray}}
\newcommand\ea{\end{eqnarray}}
\newcommand\bt{\bibitem}
\newcommand\nn{\nonumber}
\newcommand\lt{\left}
\newcommand\rt{\right}
\newcommand\pt{\partial}
\newcommand\tx{\text}
\newcommand\mc{\mathcal}
\begin{document}
\title{Probing a NED inspired Magnetically Charged Black Hole in the Hernquist Dark Matter Halo}
\author{Sohan Kumar Jha}
\email{sohan00slg@gmail.com}
\affiliation{Department of Physics, Chandernagore College, Chandannagore, Hooghly, West
Bengal, India}

\date{\today}
\begin{abstract}
\begin{center}
Abstract
\end{center}
With an intent to examine the combined effect of non-linear electrodynamics (NED) and dark matter (DM), we obtain a static and spherically symmetric solution with the black hole (BH) magnetically charged and immersed in the \h (MHDM). The position of the event horizon $r_h$ and the critical impact parameter $b_m$ are then probed to gauge the extent of influence magnetic charge $g$ and halo parameters $\a$, $\b$ have on them. A recurring outcome of our analysis with respect to different BH observables is the nullification of competing effects of charge and halo parameters, leading to observables obtaining values equal to those for a \s BH. This is also observed for $r_h$ and $b_m$. We delve into unraveling the impact of NED and DM combined on the strong gravitational lensing (GL) and its related observables, such as the angular separation, relative magnification, and the angular position of the inner, closely packed bright ring. Interestingly, we find combinations of charge and halo parameters that leave the deflection angle unchanged from the \s case, thereby leading to a situation where an MHDM BH and a \s BH become indistinguishable. Similar results are also observed for lensing observables. Finally, utilizing observations related to the angular diameter of super-massive BHs (SMBHs) $M87^*$ and $SgrA^*$ and employing the $\chi^2$ test, we extract bounds on $g$, $\a$, and $\b$ signifying the viability of our BH model as an SMBH.

\textbf{Keywords:} Dark matter, Strong gravitational lensing, Shadow, Parameter estimation, Magnetic charge, NED.
\end{abstract}
\maketitle
\section{Introduction}
BH solutions in modified theories of gravity exhibit significant deviations from general relativity (GR), providing a unique testing ground for high-energy astrophysics. These deviations can be observed in phenomena such as the BH shadow and GL. One way to modify GR is to involve NED, where the strong-field effects are incorporated by suitable modification of Maxwell's theory \c{ned1, ned2}. These modifications, in high-energy, result naturally around magnetically charged BHs or for intense electromagnetic fields \c{ned3, ned4}. Several studies have pointed out the ability of NED to circumvent the singularity problem, thereby giving rise to regular BHs and completeness of geodesics without modifying the essential thermodynamic properties \c{ned5, ned6, ned7}. In \c{ned71}, authors have illustrated that the first ever regular BH proposed, i.e., Bardeen BH, can be interpreted as a result of coupling between GR and NED. NED coupled GR has also been employed to obtain singular BH solutions \c{ned12, ned16}. There may arise two different scenarios related to the NED Lagrangian: one is where the Lagrangian depends on the electric and/or magnetic charge, such as in \c{ned1, ned71}, or is independent of charge, such as in \c{ned17}. There exist two frameworks, $F$ and $P$, that can be used to represent NED models. In the former framework, the NED Lagrangian, $\l (F)$, is gauge-invariant and a function of the Maxwell scalar $F=\f{1}{4}F\mn F\MN$, where $F\mn$ is the tensor related to the electromagnetic field. For $P$ framework, the Lagrangian $\l (F)$ is converted to $\l (P)$ through a Legendre transformation where $P=\f{1}{4}P\mn P\MN$ with $P\mn=\l_{F}F\mn$ being the auxiliary electromagnetic field. We will be using the $F$ framework for our current study. Please see \c{ned8, ned9, ned10, ned11, ned13} for studies related to different aspects of NED modified BHs.\\  
It is implausible that astrophysical BHs exist in a vacuum, as they are likely surrounded by matter fields. Dark matter (DM), a prominent candidate for these fields, has attracted significant research interest. Evidence for DM emerged from observations of elliptical and spiral galaxies \c{rubin}, with one study estimating DM comprises about $90\%$ of a galaxy's mass \c{persic}. Substantial evidence further suggests astrophysical BHs may reside within DM halos \cite{akiyamal1, akiyamal6, akiyamal12}. These findings emphasize the importance of including DM near galactic centers \c{sofue, boshkaye}. To model DM’s effect, several DM density profiles are used [\citenum{kiselev} - \citenum{rayimbaev}]. The Dehnen profile is especially versatile, with different model parameters generating a variety of DM distributions \c{dehnen, mo}. In this work, we consider the NED Lagrangian from \c{ned16}and the Hernquist distribution \c{skj} to study the combined effect of DM and NED. For further studies on DM halo-BH systems, see [\citenum{jusufi19} - \citenum{SN}].\\
Very few topics have attracted as much attention and extensive research as gravitational lensing, for it embeds the intrinsic properties of the concerned spacetime. Since the nature of spacetime governs a photon's trajectory, it carries with it the signature of the strong-field regime just outside the event horizon. Since theories of modified gravity must agree with GR in the weak field limit, it is pertinent to explore strong fields to examine divergence from GR. This makes GL such an excellent avenue in the strong-field limit. When a photon's trajectory grazes a compact astrophysical object, its path gets deflected from its original course due to the compact object's gravitational effect. Such a phenomenon is referred to as gravitational lensing, and the compact object is called a gravitational lens. Darwin, in his pioneering work \c{dar}, applied for the first time the concept of GL to the \s BH, which provided the foundation for future developments in this regard. The gravitational lens equation conceptualized in \c{vir} furthered our understanding. Bozza and others in \c{BOZZA} have developed further the analytical method that adds more tools to study GL. There exist numerous studies that explore GL to examine possible signatures of deviations from GR warranted by its modification. Please refer [\citenum{BOZZA1}-\citenum{lens13}] for the application of GL to different spactimes and see \c{id1, id2, id3} where GL is employed to find the imprint of DM. We will be using experimental observations related to the angular diameter of SMBHs $M87^*$ and $SgrA^*$ to constrain parameters \c{akiyamal1, akiyamal12, sgra}.\\      
This article is organized as follows. Section II is where we find the BH solution endowed with magnetic charge and DM. In Sections III and IV, we examine the strong GL and related observables to probe the possible impact of charge and DM halo. Section V utilizes the foundation laid in preceding sections, observational data, and the $\chi^{2}$ test to extract constraints on charge and halo parameters. Section VI concludes this article by providing a brief overview of the obtained results. We have used $G=c=M=1$ unless stated otherwise.
\section{magnetically charged BH in a Hernquist DM halo}
We, in this section, aim to obtain a new BH configuration where a magnetically charged BH arising out of NED is immersed in the \h (MHDM BH). The action for the composite system is given by 
\begin{widetext}
\begin{align}\label{action}
S=\int d^4x\sqrt{-g}\bigg[\frac{R}{2\kappa}-\f{2\mathcal{L}(F)}{\kappa}+\mathcal{L}_{dm}\bigg],
\end{align}
\end{widetext}
where $R$ is the Ricci scalar, $\mc{L}(F)$ is the NED Lagrangian density which is a function of the electromagnetic field invariant $F$ defined by \\
\be
F=\f{1}{4}F\mn F\MN,
\ee
with $F\mn=\partial_{\mu}A_{\nu}-\partial_{\nu}A_{\mu}$ being the field strength of the NED four-vector potential $A_{\mu}$ and $\mc{L}_{dm}$ is the Lagrangian density of the \h. The Lagrangian for NED $\mc{L}(F)$ tends to $F$ in the Maxwell limit. We obtain relevant field equations by varying the action (\r{action}) with respect to $g\MN$. They are 
\ba
R_{\mu \nu }-\frac{1}{2}g_{\mu \nu }R=\kappa  T^{\tx{H}}_{\mu\nu}+2\lt(\f{\pt \mc{L}}{\pt F}F_{\mu \lambda}F_{\nu}^{\lambda}-g_{\mu\nu}\mc{L}(F)\rt).
\label{fe}
\ea
The following are, respectively, the dynamic equation and the Bianchi identity obeyed by the field strength tensor:
\ba
\nabla_\mu\left(\frac{\partial \mathcal{L}(F)}{\partial F} F^{\nu \mu}\right)&=&0,\\
\nabla_\mu\left(* F^{\nu \mu}\right)&=&0.\label{fmu}
\ea
With the above equations at our disposal, we now move on to the ingredients required to solve them. The density profile for the \h is \c{mo}
\be
\rho_{H}(r)=\a \lt(\f{r}{\b}\rt)^{-1}\lt[1+\f{r}{\b}\rt]^{-3},\label{hd}
\ee
where $\a$ is the chracteristic density and $\b$ is the characteristic radius of the halo.The metric for the pure \h is given by \c{skj}
\begin{equation}
ds^2 = -\mathcal{A}_1(r) dt^2 + \mathcal{A}_1(r)^{-1} dr^2 + r^2 (d\theta^2 + \sin^2 \theta d\phi^2),
\label{ldm}
\end{equation}
with 
\be
\mathcal{A}_1(r)=1-\f{4\pi \a \b^3}{r+\b}.
\ee  
The energy-momentum tensor of the DM halo is $T_{\mu}^{\nu\,(H)}=diag(-\rho, p_r,\, p_\theta,\, p_\phi)$ with the energy density, radial, and tangential pressures given by 
\be
-\rho=p_r=-\frac{\alpha  \beta ^4}{2 r^2 (\beta +r)^2} \quad \tx{and} \quad p_\theta=p_\phi=\frac{\alpha  \beta ^4}{2 r^2 (\beta +r)^2}. \label{density}
\ee 
Next, we consider the following form for the field strength \c{ned16}
\begin{equation}
F_{\mu \nu}=\left(\delta_\mu^\theta \delta_\nu^{\phi}-\delta_\nu^\theta \delta_\mu^{\phi}\right) B(r, \theta)=\left(\delta_\mu^\theta \delta_\nu^{\phi}-\delta_\nu^\theta \delta_\mu^{\phi}\right) g(r) \sin \theta.\label{fnu}
\end{equation}
If we employ Eq. (\r{fmu}), we get $g(r)=\tx{Constant}=g$. Here, $g$ signifies the magnetic monopole charge. Eq. (\r{fnu}) leads us to
\be
F_{\theta\phi}=-F_{\phi\theta}=g\sin \theta \quad \tx{and} \quad F=\f{g^2}{2r^4}.
\ee
Following is the NED Lagrangian density considered in this manuscript: 
\begin{equation}
\mathcal{L}(F)=\frac{2 \sqrt{g} F^{5 / 4}}{s(\sqrt{2}+2 g \sqrt{F})^{3 / 2}},
\end{equation}
where $s$ is a constant to be fixed later.\\
We assume the following ansatz to obtain NED inspired static, spherically symmetric BH solution immersed in the \h:
\be
ds^2=-A(r)dt^2+B(r)dr^2+r^2 d\theta^2+r^2 \sin^2\theta d\phi^2.
\label{trial}
\ee 
Having requisite expressions at our disposal, we now move to solving field equations, which are given by
\ba
&&-\frac{r B'(r)+(B(r)-1) B(r)}{r^2 B(r)^2}=-\kappa\rho-2\mc{L},\label{00}\\
&&\frac{r A'(r)-A(r) B(r)+A(r)}{r^2 A(r) B(r)}=\kappa p_r-2\mc{L},\label{11}\\\nn
&&\frac{-r B(r) A'(r)^2+A(r) \left(2 B(r) \left(r A''(r)+A'(r)\right)-r A'(r) B'(r)\right)-2 A(r)^2 B'(r)}{4 r A(r)^2 B(r)^2}=\kappa p_{\theta}-2\lt(\f{r}{2}\f{\pt \mc{L}}{\pt r}+\mc{L}\rt),\label{22}
\ea
Eq. (\r{00}) together with eq. (\r{11}) leads to the relation between $\ar$ and $\br$ as 
\be
B(r) A'(r)+A(r) B'(r)=0,
\ee
yielding $\br=\f{1}{\ar}$. This along with Eq. (\r{00}) leads us to the desired ansatz
\be
ds^2=-A(r)dt^2+A(r)^{-1}dr^2+r^2 d\theta^2+r^2 \sin^2\theta d\phi^2,\label{final}
\ee
with
\be
\ar=1-\f{2M}{\sqrt{r^2+g^2}}-\f{4\pi \a \b^3}{r+\b}.
\ee  
The metric, in the limit $g \rightarrow 0$ and $\a(\b) \rightarrow 0$, reverts to the \s BH. The metric has one conspicuous singularity at $r=r_h$ where $A(r_h)=0$. This marks the position of the event horizon. However, the metric components show no divergence at $r=0$, leading one to believe that the obtained solution may be a regular one endowed with completeness of geodesics. This air over the singularities can be cleared with the help of scalar invariants given as follows:
\ba\nn
&&\tx{Ricci Scalar}=R=\frac{2}{r^2} \left(\frac{g^2 M \left(2 g^2-r^2\right)}{\left(g^2+r^2\right)^{5/2}}+\frac{4 \pi  \alpha  \beta ^5}{(\beta +r)^3}\right),\\\nn
&&\tx{Ricci squared}=R\mn R\MN\\\nn
&&=\frac{1}{2 r^4}\left(\frac{32 \pi  \alpha  \beta ^4 g^2 M \left(2 \beta  \left(g^2+r^2\right)+2 g^2 r+5 r^3\right)}{\left(g^2+r^2\right)^{5/2} (\beta +r)^3}+\frac{4 g^4 M^2 \left(4 g^4+8 g^2
   r^2+13 r^4\right)}{\left(g^2+r^2\right)^5}+\frac{64 \pi ^2 \alpha ^2 \beta ^8 \left(\beta ^2+2 r^2+2 \beta  r\right)}{(\beta +r)^6}\right),\\\nn
&&\tx{Kretschmann Scalar}=K\\\nn
&&=\left(\frac{8 \pi  \alpha  \beta ^3}{(\beta +r)^3}-\frac{2 M \left(g^2-2 r^2\right)}{\left(g^2+r^2\right)^{5/2}}\right)^2+\frac{4}{r^4} \left(r^2 \left(\frac{2 M
   r}{\left(g^2+r^2\right)^{3/2}}+\frac{4 \pi  \alpha  \beta ^3}{(\beta +r)^2}\right)^2+\left(\frac{2 M}{\sqrt{g^2+r^2}}+\frac{4 \pi  \alpha  \beta ^3}{\beta
   +r}\right)^2\right).\\
\label{inv}
\ea 
The above expressions make it clear that the metric (\r{final}) is not devoid of singularity at $r=0$, which is, in this case, an essential one that cannot be circumvented with the help of any coordinate transformation. The absence of singularity at $r=r_h$ in the above expressions makes it avoidable through a suitable coordinate transformation. We now explore the dependence of the event horizon on the magnetic charge $g$ and the halo parameters $\a$ and $\b$.
\begin{figure}[H]
\begin{center}
\begin{tabular}{cc}
\includegraphics[width=0.4\columnwidth]{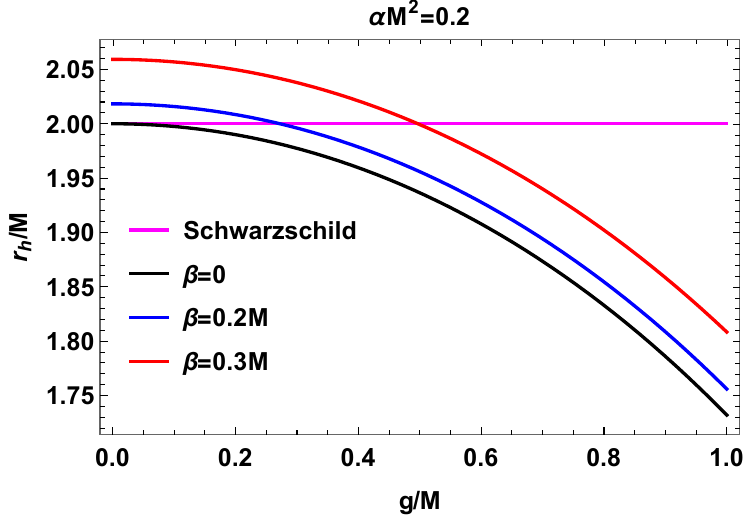}&
\includegraphics[width=0.4\columnwidth]{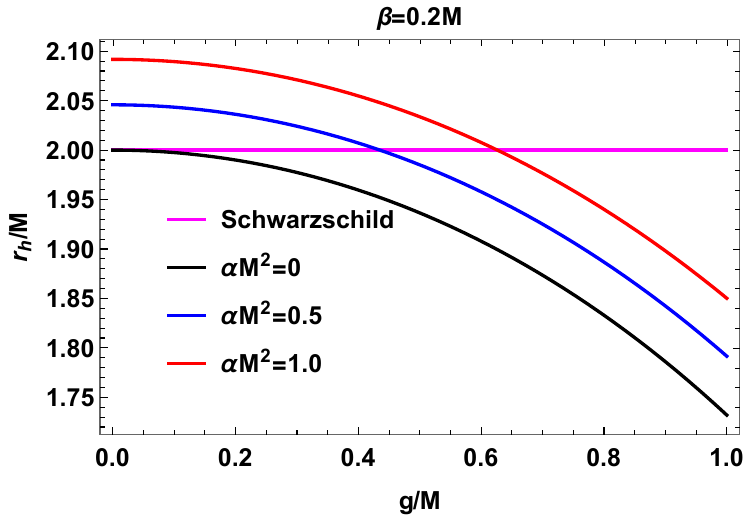}
\end{tabular}
\caption{Variation of event horizon with $g$ for different values of $\b$ (left panel) and different values of $\a$ (right panel). }\label{rh1}
\end{center}
\end{figure}
\begin{figure}[H]
\begin{center}
\begin{tabular}{cc}
\includegraphics[width=0.4\columnwidth]{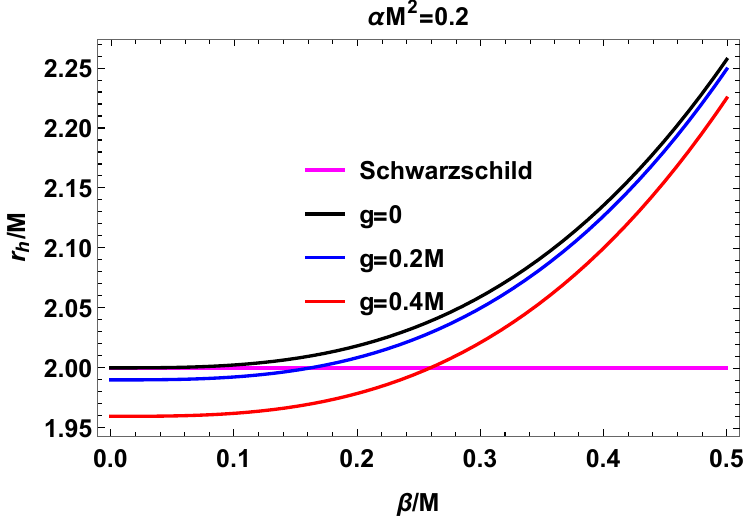}&
\includegraphics[width=0.4\columnwidth]{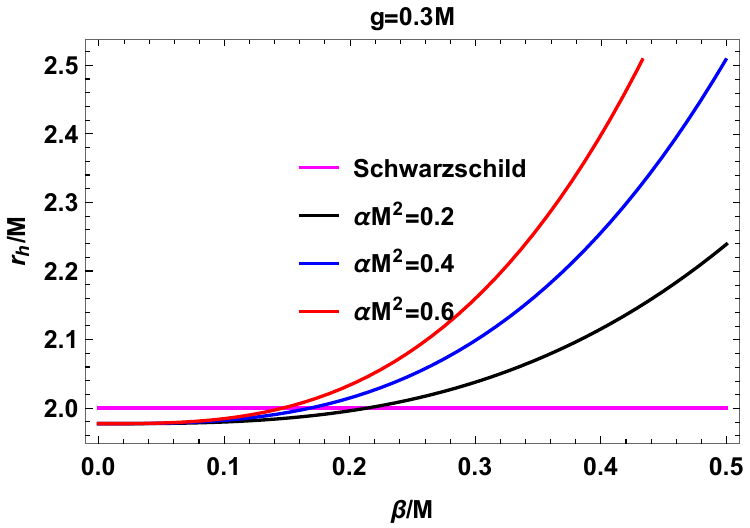}
\end{tabular}
\caption{Variation of event horizon with $\b$ for different values of $g$ (left panel) and different values of $\a$ (right panel). }\label{rh2}
\end{center}
\end{figure}
\begin{figure}[H]
\begin{center}
\begin{tabular}{cc}
\includegraphics[width=0.4\columnwidth]{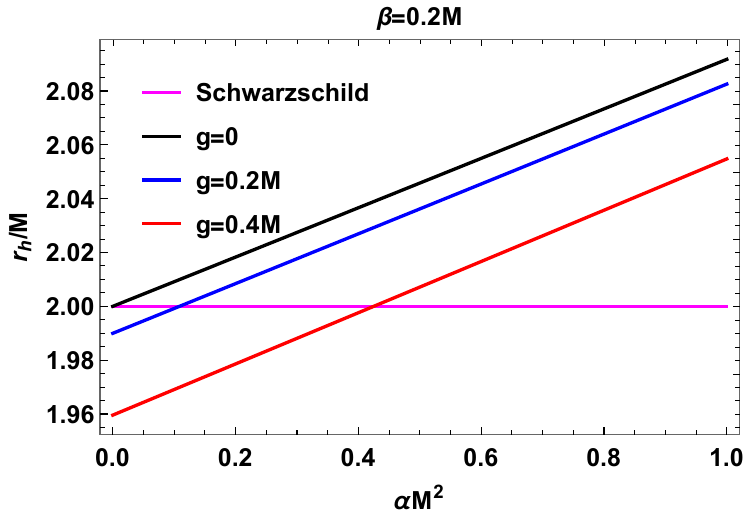}&
\includegraphics[width=0.4\columnwidth]{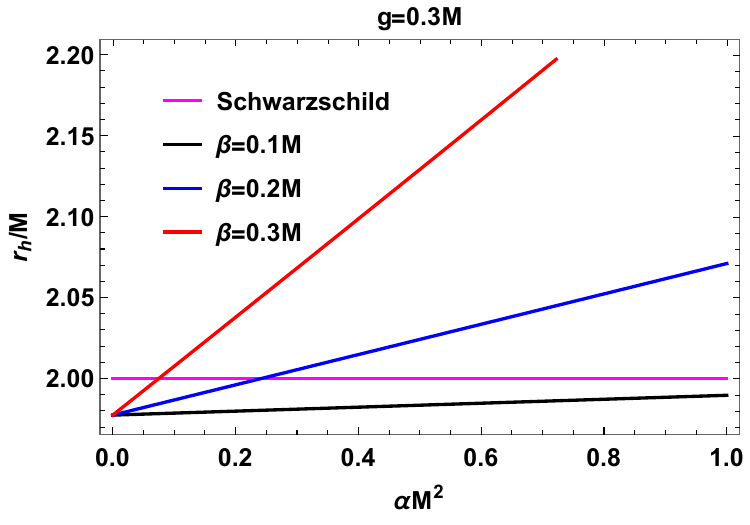}
\end{tabular}
\caption{Variation of event horizon with $\a$ for different values of $g$ (left panel) and different values of $\b$ (right panel). }\label{rh3}
\end{center}
\end{figure}
Fig. (\r{rh1}), (\r{rh2}), and (\r{rh3}) clearly demonstrate the visible effect of magnetic charge and DM halo on the event horizon. While the magnetic charge adversely impacts the event horizon, the DM halo parameters augment its value. For non-null values of the magnetic charge, there exist critical values of free parameters that mark the transition from a lesser value of the event horizon than the \s case to greater values. For example, at $\a M^2=0.15,\,\b=0.15M$ we have a critical value of magnetic charge $g_{ch}=0.154M$ where we have a smaller value of event horizon than the \s case when $g>g_{ch}$ and reverse is the case when $g<g_c$. At $g=g_{ch}$, we have $r_h=2M$, making an MHDM BH and a \s BH indistinguishable when it comes to the position of the event horizon. Similarly, at $g=0.2M$ and $\a M^2=0.2$, we have $\b_{ch}=0.1625M$ and at $g=\b=0.1M$, we have $\a_{ch} M^2=0.21$. However, unlike the magnetic charge case, we have smaller values of $r_h$ when $\a(\b)<\a_{ch}(\b_{ch})$. Next, strong GL will be explored to gauge the interplay between NED and DM halo.
\section{strong gravitational lensing by an mhdm BH}
This section serves two purposes: one is to probe the combined effect of NED and DM on GL, and the second is to lay the foundation that will eventually help us extract parameter values concordant with experimental observations. Following the treatment expounded in \cite{BOZZA, BOZZA1, BOZZA2}, we rewrite the ansatz (\r{final}), confining only to the equatorial plane, as
\begin{equation}
d\tilde{s}^{2}=(2M)^{-2}ds^{2}=-F(x) dt^{2}+F(x)^{-1} dx^{2}+H(x) d \phi^{2}, \label{final1}
\end{equation}
where $x=\f{r}{2M}$, $\tb=\f{\b}{2M}$, $\ta=4M^2\a$, $\tg=\f{g}{2M}$, and
\be
\fx= 1-\f{1}{\sqrt{x^2+\tg^2}}-\f{4\pi \ta \tb^3}{r+\tb}\quad \tx{and} \quad \hx=x^2.
\ee  
We next write the Lagrangian for the metric (\r{final1}) that will lead us to the position of the unstable photon orbit that is responsible for the observed bright ring. Following is the required Lagrangian:  
\be
\l=\f{1}{2}\lt(-\fx \dot{t}^2+\f{\dot{x}^2}{\fx}+\hx \dot{\phi}^2\rt),\label{l}
\ee
where $\dot{t}=\f{dt}{d\lambda}$ and $\dot{\phi}=\f{d\phi}{d\lambda}$, with $\lambda$ representing the affine parameter. The absence of $t$ and $\phi$ in the Lagrangian leads us to two conserved quantities that are associated with null geodesics: one is the energy $E$, and another is the angular momentum $L$ given by
\be
E=-\f{d\l}{d\dot{t}}=\fx \dot{t} \quad \tx{and} \quad L=\f{d\l}{d\dot{\phi}}=\hx \dot{\phi}.
\ee
The above expressions, along with the fact that we have $d\tilde{s}=0$ for null geodesics, eventually lead us to the following differential equation:
\be
\dot{x}^2=E^2-\f{L^2\fx}{\hx}=E^2-V(x),
\ee
where $V(x)=\f{L^2\fx}{\hx}$ is the potential for the photon trajectory. We have two conditions being imposed on the potential for circular orbits with radius $x_m$: $V(x_m)=E^2$ and $\f{dV}{dx}|_{x=x_m}=0$. The second condition leads us to the following equation:
\be
H'(x)\fx=F'(x)\hx,
\ee
where prime means differentiation with respect to the argument $x$. With each impact parameter $b$, there associates a minimum distance $x_0$ that follows the condition:
\be
\f{dx}{d\phi}=0 \quad \Rightarrow \quad b=\f{L}{E}=\sqrt{\f{H(x_0)}{F(x_0)}}.
\ee
The limiting value of $b$, $b_m$, is reached when the minimum distance coincides with the photon radius $x_m$. Any photon that travels with an impact parameter less than the limiting value gets annihilated by the BH. This limiting value of the impact parameter serves as the shadow radius for an asymptotic observer. The above equation for the MHDM metric can not be solved analytically, and as such, we have resorted to a numerical method. Fig. (\r{rs1}), (\r{rs2}), and (\r{rs3}) demonstrate how variation of magnetic charge or halo parameters affects the shadow radius.
\begin{figure}[H]
\begin{center}
\begin{tabular}{cc}
\includegraphics[width=0.4\columnwidth]{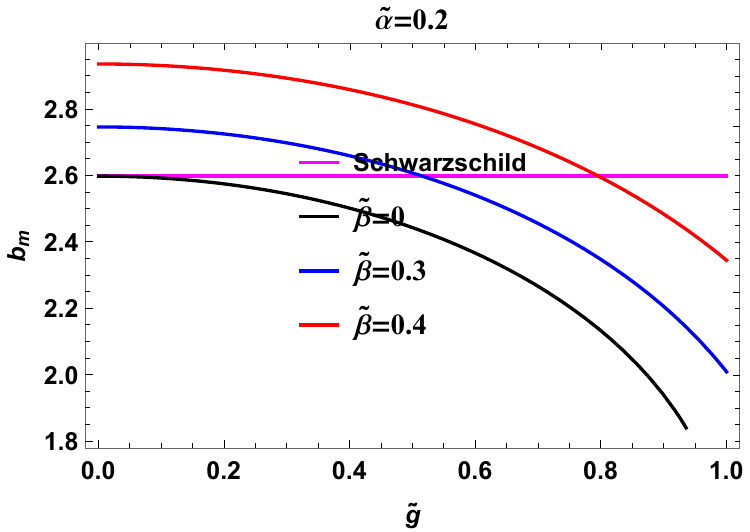}&
\includegraphics[width=0.4\columnwidth]{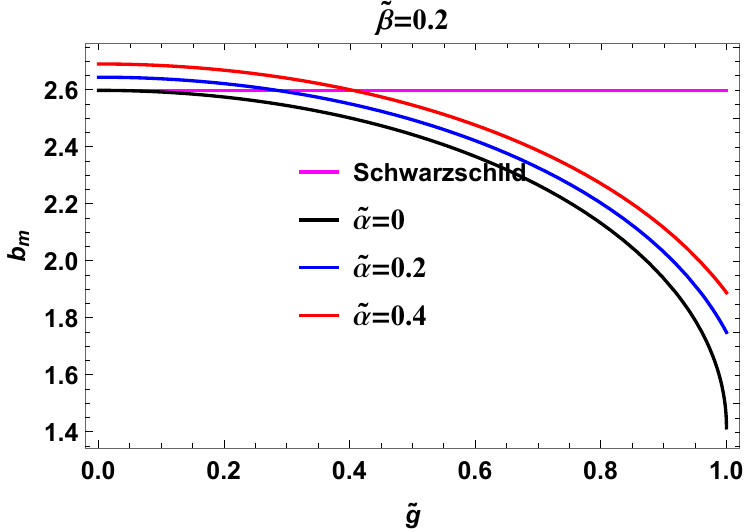}
\end{tabular}
\caption{Variation of shadow radius with $\tg$ for different values of $\tb$ (left panel) and different values of $\ta$ (right panel). }\label{rs1}
\end{center}
\end{figure}
\begin{figure}[H]
\begin{center}
\begin{tabular}{cc}
\includegraphics[width=0.4\columnwidth]{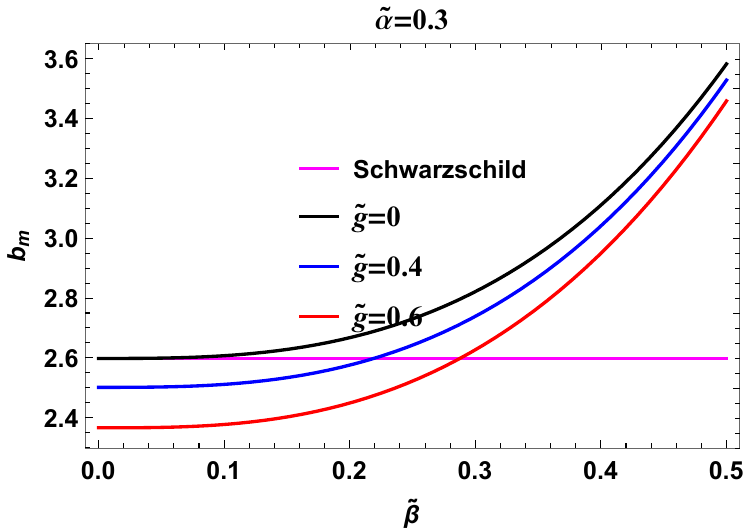}&
\includegraphics[width=0.4\columnwidth]{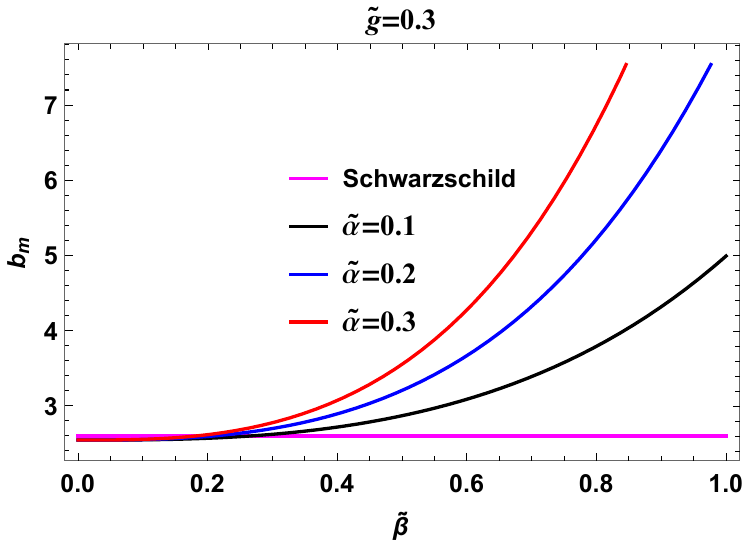}
\end{tabular}
\caption{Variation of shadow radius with $\tb$ for different values of $\tg$ (left panel) and different values of $\ta$ (right panel). }\label{rs2}
\end{center}
\end{figure}
\begin{figure}[H]
\begin{center}
\begin{tabular}{cc}
\includegraphics[width=0.4\columnwidth]{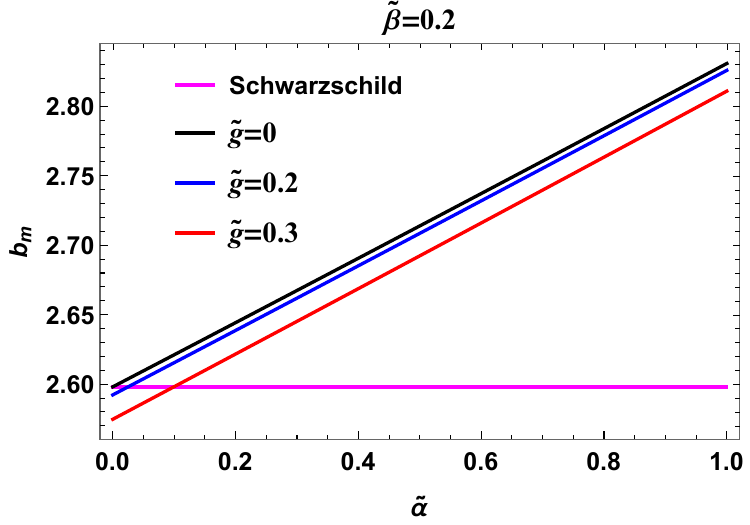}&
\includegraphics[width=0.4\columnwidth]{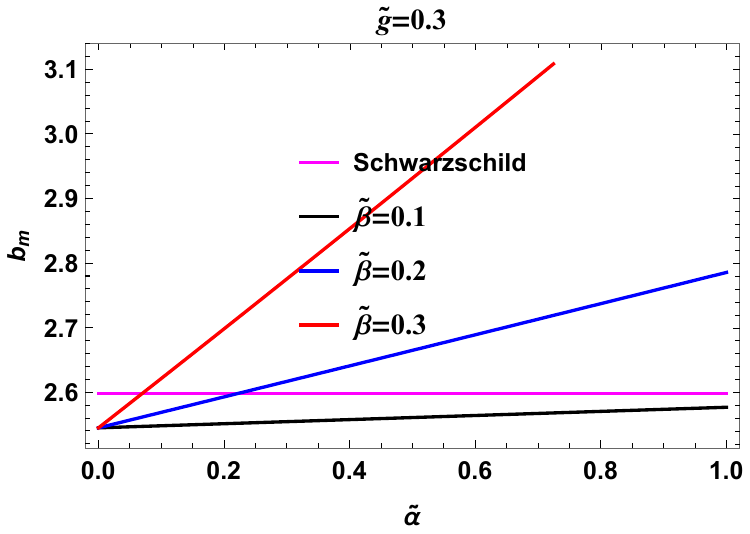}
\end{tabular}
\caption{Variation of shadow radius with $\ta$ for different values of $\tg$ (left panel) and different values of $\tb$ (right panel). }\label{rs3}
\end{center}
\end{figure}
The shadow radius shows similar dependence on charge and halo parameters as that of the event horizon. Like the horizon case, here too we observe critical values of charge $\tg_{cm}$ and halo parameters $\ta_{cm}$ and $\tb_{cm}$. These values mark the point where an MHDM BH and a \s BH can not be distinguished based on their shadow size. Some of the examples of such critical values are: at $\ta=0.3,\, \tb=0.2$ we have $\tg_{cm}=0.3495$, at $\tg=0.2$, $\tb=0.15$ we have $\ta_{cm}=0.2283$, and we have $\tb_{cm}=0.2075$ at $\tg=0.3$, $\ta=0.2$. Following articles \c{vir}, we write the deflection angle as
\begin{equation}
\alpha_D\left(x_{0}\right)=I\left(x_{0}\right)-\pi,\label{def}
\end{equation}
where
\begin{equation}
I\left(x_{0}\right)=\int_{x_{0}}^{\infty}\frac{2}{\sqrt{F(x)H(x)}
\sqrt{\frac{F(x_{0})H(x)}{H(x_{0})F(x)}-1}}dx.\label{io}
\end{equation}
The above expression diverges when the minimum distance coincides with the photon radius, i.e., at $x_0=x_m$. To circumvent this issue, we introduce the variable $z=1-\f{x_0}{x}$. Following the prescription elucidated in \c{BOZZA}, we eventually arrive at the following expression of the deflection angle given by
\begin{eqnarray}
&& \gamma_D(b)=-\overline{a} \log \left( \frac{b}{b_m} -1
\right) +\overline{b}+O(b-b_m) ,\label{alphab}\\\nn%
\text{where}\\
&& \overline{a}=\frac{a}{2}=\frac{\mathcal{R}(0,x_m)}{2\sqrt{a_2(x_m)}}, \label{afinal}\\%
&& \overline{b}=-\pi+\bar{b}_\mathcal{R}+\overline{a}\log{\frac{2H^2(x_m)a_2(x_m)}{F(x_m)x_m^4}}\label{bfinal},\\\nn
\text{with}\\\nn
&& \mathcal{R}(z,x_m)=\frac{2x^{2}\sqrt{H(x_0)}}{x_{0}H(x)},\\\nn
&&a_{2}(x_0) = \frac{1}{2}\left[\frac{\lt(2x_{0}H(x_{0})-2x_{0}^{2}H^{\prime}(x_{0})\rt)\lt(H^\prime(x_0) F(x_0)-F^\prime(x_0) H(x_0)\rt)}{H^{2}(x_{0})}+\frac{x_{0}}{H(x_{0})}\left(H^{\prime\prime}(x_{0})F(x_{0})-F^{\prime\prime}(x_{0})F(x_{0})\right)\right],\\\nn
&&g(z,x_0)=\mathcal{R}(z,x_0)f(z,x_0)-\mathcal{R}(0,x_m)f_0(z,x_0),\\\nn
&& I_\mathcal{R}(x_0)=\int\limits_0^1 g(z,x_m) dz+O(x_0-x_m) \quad \text{and} \quad \bar{b}_\mathcal{R}= I_\mathcal{R}(x_m).
\end{eqnarray}
In the absence of the magnetic charge and DM halo, Eq. (\r{afinal}) and (\r{bfinal}) yield $\ab=1$ and $\bb=-0.40023$, their values for the \s BH \c{BOZZA}. Fig. (\r{ab1}), (\r{ab2}), and (\r{ab3}) showcase qualitatively the impact of magnetic charge and DM halo on the lensing coefficient $\ab$ and Fig. (\r{bb1}), (\r{bb2}), and (\r{bb3}) demonstrate those for $\bb$.  
\begin{figure}[H]
\begin{center}
\begin{tabular}{cc}
\includegraphics[width=0.4\columnwidth]{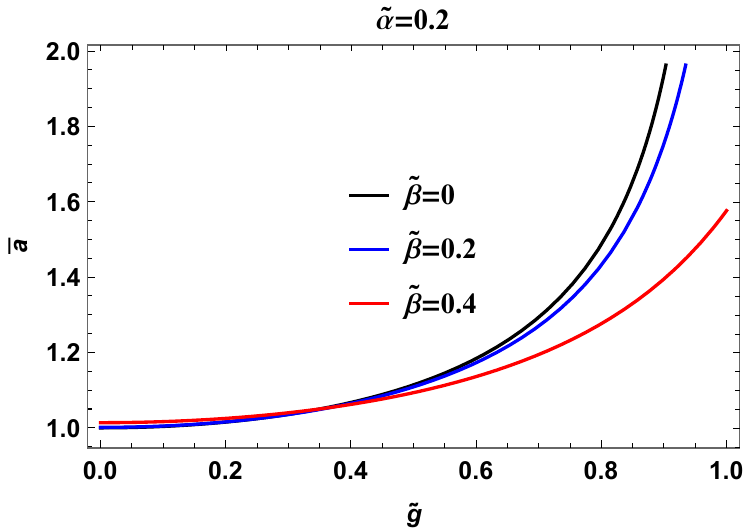}&
\includegraphics[width=0.4\columnwidth]{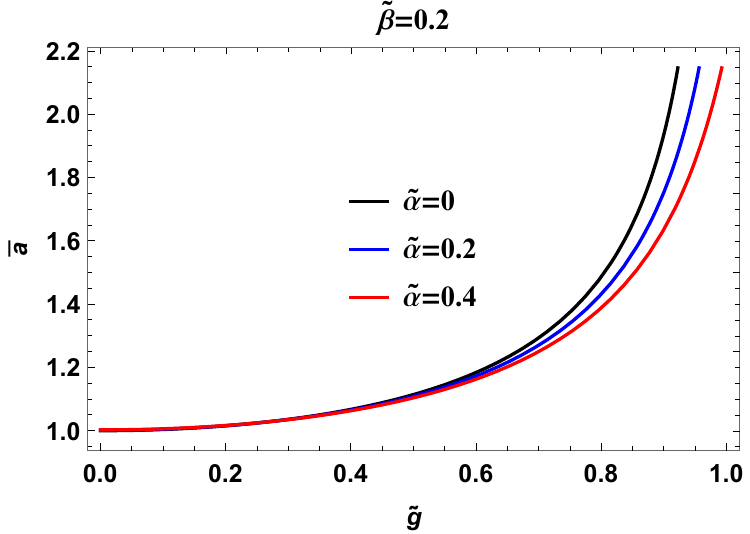}
\end{tabular}
\caption{Variation of lensing coefficient $\ab$ with $\tg$ for different values of $\tb$ (left panel) and different values of $\ta$ (right panel). }\label{ab1}
\end{center}
\end{figure}
\begin{figure}[H]
\begin{center}
\begin{tabular}{cc}
\includegraphics[width=0.4\columnwidth]{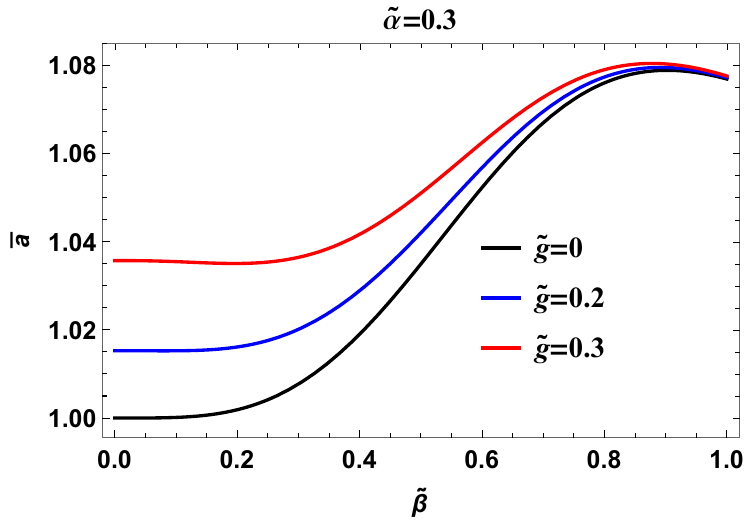}&
\includegraphics[width=0.4\columnwidth]{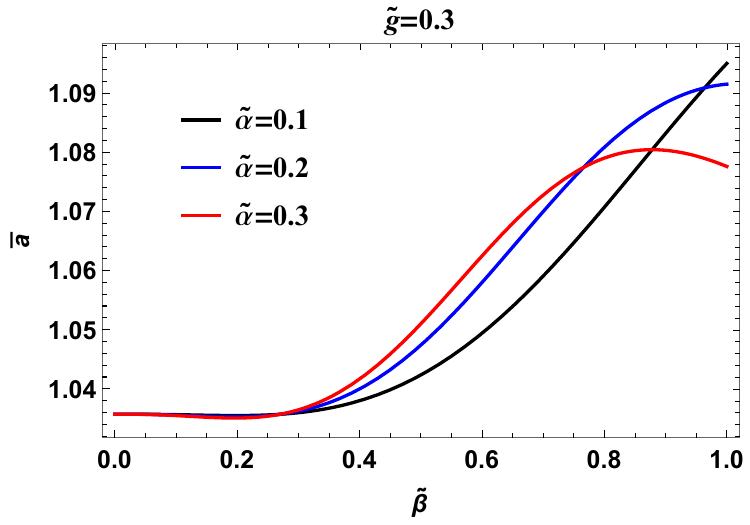}
\end{tabular}
\caption{Variation of lensing coefficient $\ab$ with $\tb$ for different values of $\tg$ (left panel) and different values of $\ta$ (right panel). }\label{ab2}
\end{center}
\end{figure}
\begin{figure}[H]
\begin{center}
\begin{tabular}{cc}
\includegraphics[width=0.4\columnwidth]{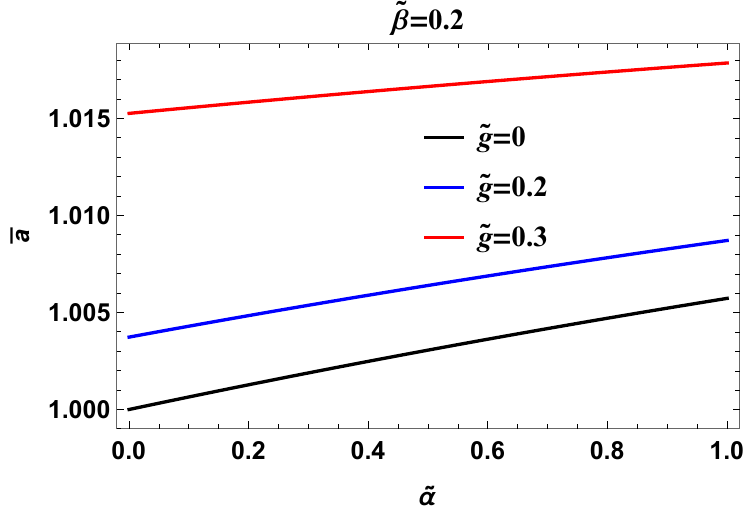}&
\includegraphics[width=0.4\columnwidth]{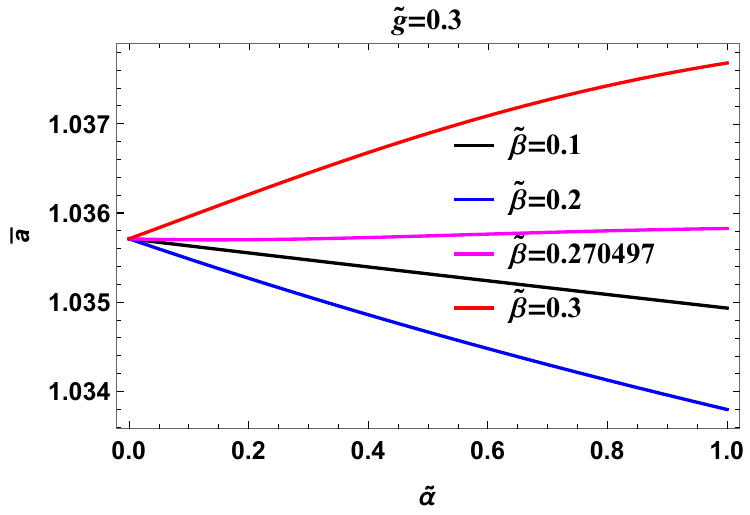}
\end{tabular}
\caption{Variation of lensing coefficient $\ab$ with $\ta$ for different values of $\tg$ (left panel) and different values of $\tb$ (right panel). }\label{ab3}
\end{center}
\end{figure}
For an MHDM BH, the lensing coefficient $\ab$ is always greater than that for a \s BH. While the magnetic charge has a favourable impact on the coefficient for any value of DM parameters, the effect of $\ta$ on $\ab$ for a fixed value of $\tg$ depends on the halo parameter $\tb$. There exists a value of $\tb\,(\tb_{ca})$ that nullifies the DM effect on $\ab$. The lensing coefficient increases with $\ta$ when $\tb\,>\,\tb_{ca} $ and reverse is the case when $\tb\,<\,\tb_{ca}$. For example, when $\tg=0.3$, we have $\tb_{ca}=0.270497$ and its value at $\tg=0$ is $0.00111479$. Thus, a \s BH in the \h behaves exactly like a \s BH when it comes to their lensing coefficient $\ab$ at $\tb=0.00111479$. This coefficient displays a local minimum and maximum when varied against $\tb$.   
\begin{figure}[H]
\begin{center}
\begin{tabular}{cc}
\includegraphics[width=0.4\columnwidth]{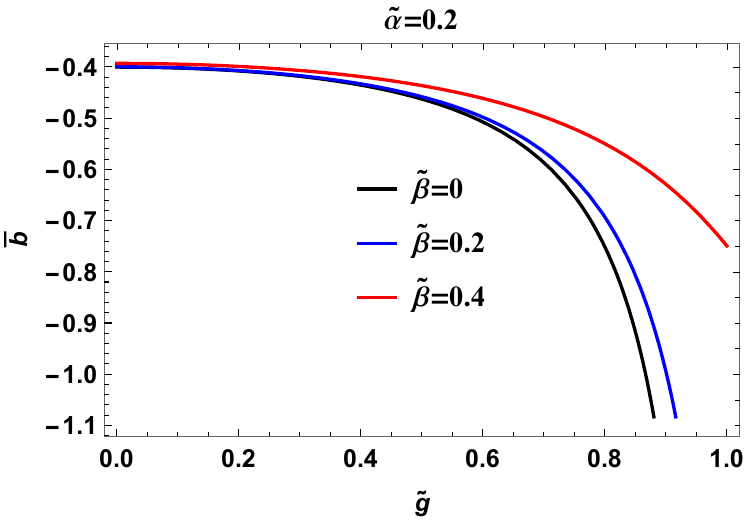}&
\includegraphics[width=0.4\columnwidth]{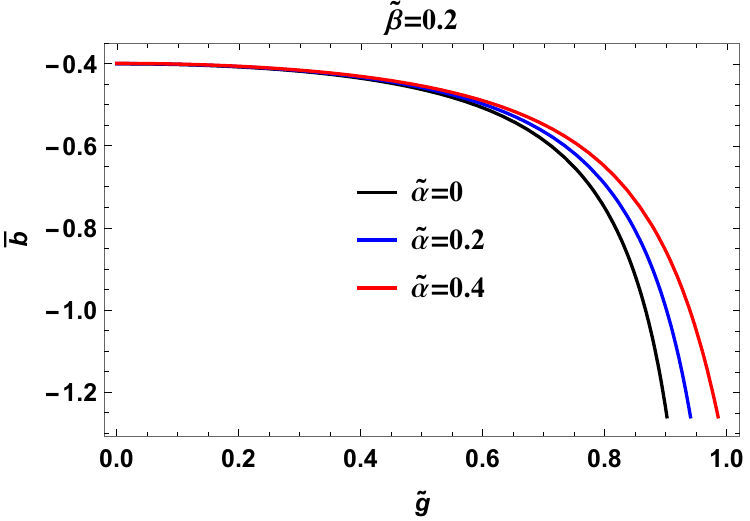}
\end{tabular}
\caption{Variation of lensing coefficient $\bb$ with $\tg$ for different values of $\tb$ (left panel) and different values of $\ta$ (right panel). }\label{bb1}
\end{center}
\end{figure}
\begin{figure}[H]
\begin{center}
\begin{tabular}{cc}
\includegraphics[width=0.4\columnwidth]{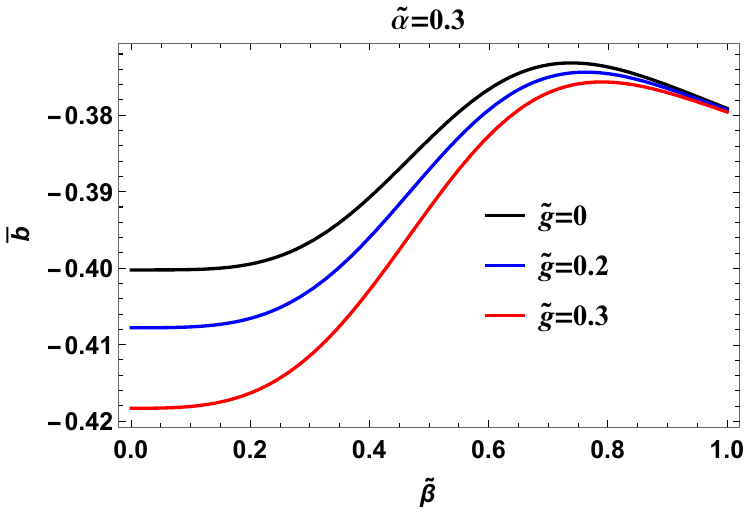}&
\includegraphics[width=0.4\columnwidth]{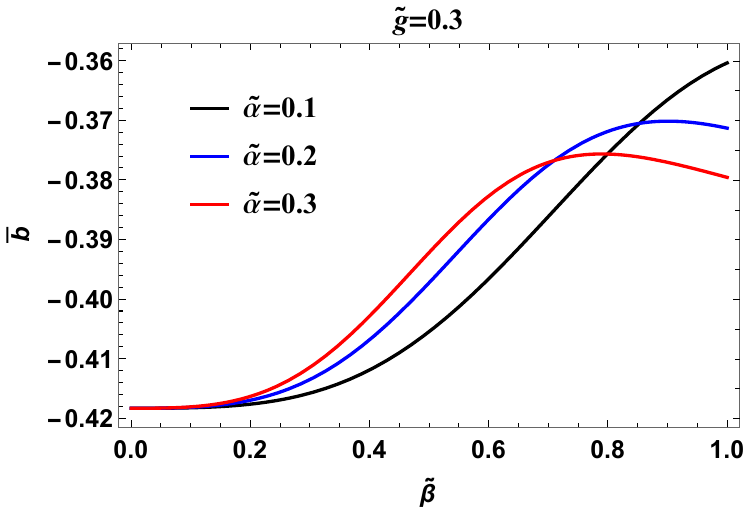}
\end{tabular}
\caption{Variation of lensing coefficient $\bb$ with $\tb$ for different values of $\tg$ (left panel) and different values of $\ta$ (right panel). }\label{bb2}
\end{center}
\end{figure}
\begin{figure}[H]
\begin{center}
\begin{tabular}{cc}
\includegraphics[width=0.4\columnwidth]{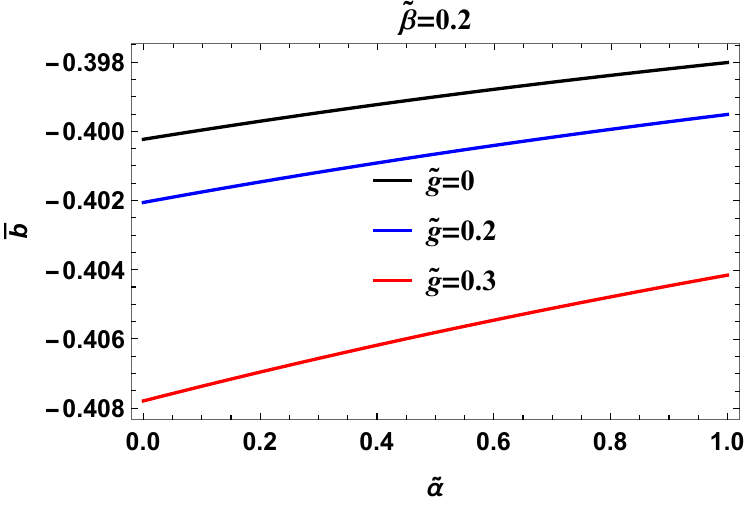}&
\includegraphics[width=0.4\columnwidth]{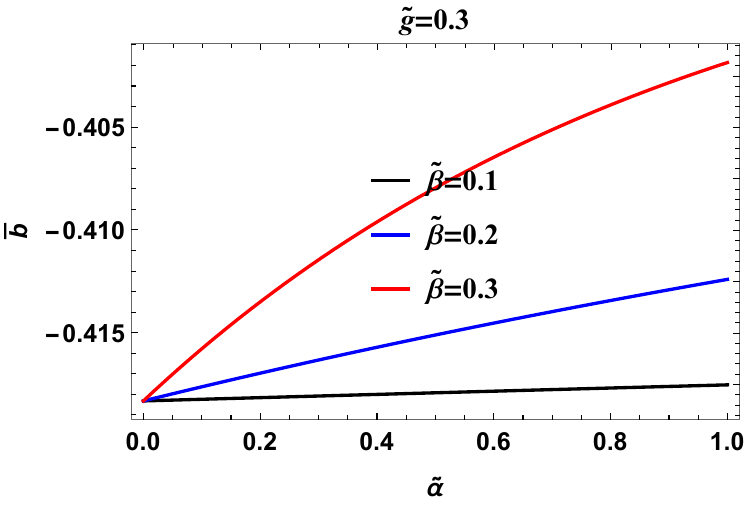}
\end{tabular}
\caption{Variation of lensing coefficient $\bb$ with $\ta$ for different values of $\tg$ (left panel) and different values of $\tb$ (right panel). }\label{bb3}
\end{center}
\end{figure}   
Fig. (\r{bb1}) displays an adverse effect of $\tg$ on the lensing coefficient $\bb$ and Fig. (\r{bb3}) demonstrates a fovourable impact of $\ta$ on $\bb$. Similar to the lensing coefficient $\ab$ case, here too we observe local minima and maxima of $\bb$ when varied against $\tb$. Having explored the qualitative variation of the lensing coefficients against charge and halo parameters, we now demonstrate the variation of the deflection angle with the impact parameter $b$ for different scenarios.
 \begin{figure}[H]
\begin{center}
\begin{tabular}{lcl}
\includegraphics[width=0.33\columnwidth]{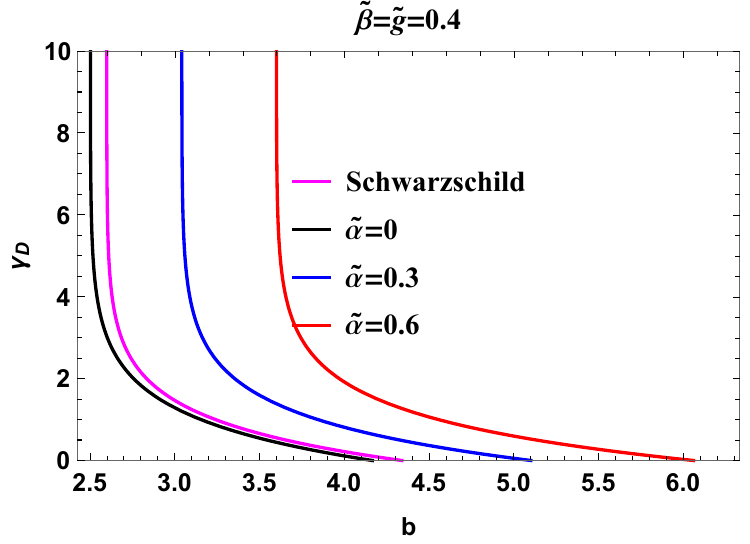}&
\includegraphics[width=0.33\columnwidth]{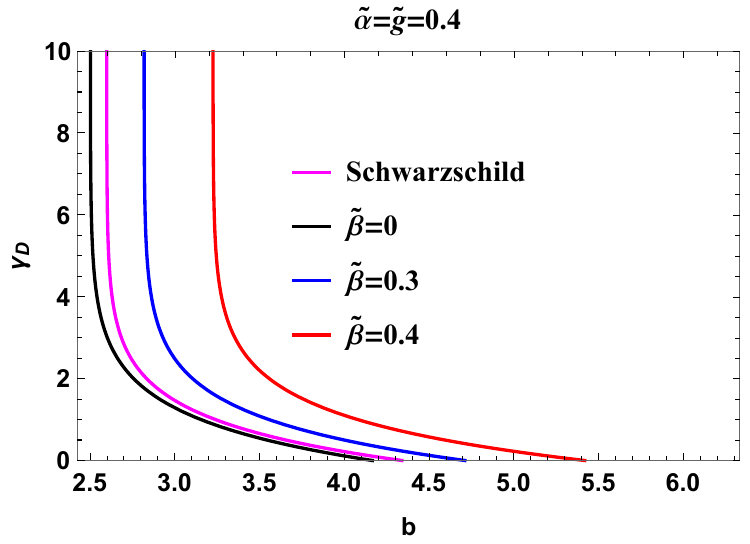}&
\includegraphics[width=0.33\columnwidth]{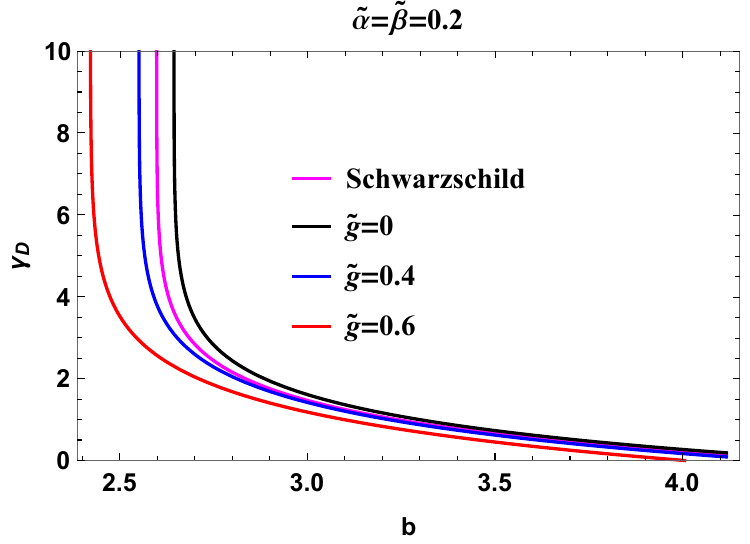}
\end{tabular}
\caption{Variation of deflection angle $\gamma_D$ with respect to the impact parameter $b$. }\label{gam}
\end{center}
\end{figure}
Fig. (\r{gam}) clearly displays visible effects of magnetic charge and halo parameters on the deflection angle. The magnetic charge adversely impacts the deflection angle, whereas the deflection angle increases with increasing halo parameters. Interestingly, Fig. (\r{gam}) points towards the existence of some combinations of $(\tg, \ta, \tb)$ that cause an MHDM BH to deflect a light ray by the same angle as that of a \s BH. Our analysis indeed reveals such combinations. Few such combinations are $(0.2,0.2,0.157153)$, $(0.1,0.18842,0.1)$, and $(0.0727999,0.1,0.1)$. A light ray will experience the same deflection when it passes by an MHDM BH with these combinations as that of a \s BH. Such cases arise due to the nullification of the augmenting effect of the DM halo by the diminishing impact of the magnetic charge. To elaborate further on the consequences of having a magnetically charged BH in the Hernquist DM halo, we will explore observables in connection with strong GL.
\section{observables in strong gravitational lensing}    
Once a photon is trapped in the unstable photon orbit, it loops around the BH several times before getting swallowed by the BH or being detected by an asymptotic observer as a result of perturbation. Those detected by the asymptotic observer form a bright ring whose interior, being black, is called the shadow of the BH. There is, however, not a single bright ring but several closely packed bright rings that form a thin bright region around the shadow. Single-loop photons are responsible for the outermost bright ring. The magnification of rings decreases exponentially as photons loop more and more around the central object. We utilize three lensing observables that will serve our purpose of gauging the impact of the interplay between NED and DM halo. Following articles \c{BOZZA, BOZZA1} and arguments thereof, expressions for the lensing observables are given by
\begin{eqnarray}
&&\theta_\infty = \frac{b_m}{D_{OL}},\quad s= \theta _1-\theta _\infty = \theta_\infty \;\; e^{\frac{\bar{b}-2\pi}{\bar{a}}},\quad \text{and \quad }r_{\text{mag}}= 2.5 \log(r) = \frac{5\pi}{\bar{a}\ln 10}\label{s} \label{obs}\\\nn
\text{where}\\
&&r = \frac{\mu_1}{\sum{_{n=2}^\infty}\mu_n } = e^{\frac{2 \pi}{\bar{a}}}.
\end{eqnarray}
Here, $D_{OL}$ represents the distance between the Earth and the BH, $\theta_\infty$ signifies the angular position of the closely packed inner bright rings resolved as one, $s$ implies the angular separation between the outermost and inner rings $\theta _1$ being the angular position of the outermost ring, $r_{\text{mag}}$ denotes the relative magnification of the outermost ring with respect to rest, and $r$ gives the ratio of fluxes from the outermost ring and the rest. With astronomical observations rendering these observables at our disposal, we can easily extract values of lensing coefficients $\ab$ and $\bb$, shedding light on the nature of MHDM BHs. SMBHs of interest for our current exposition are $M87^*$ and $SgrA^*$ whose mass $(M)$, distance from the Earth $D$, and angular diameter $\theta_d\,(=2\theta_\infty)$ are \c{akiyamal1, akiyamal12, sgra}:
\ba\nn
&&M87^*: \quad M=(6.5\,\pm\,0.7)\,\times 10^9\,M_{\odot},\quad D=(16.8\,\pm\,0.8)\, Mpc, \quad \tx{and}\quad \theta_{d}=42\,\pm\,3 \mu as,\\\nn  
&&SgrA^*: \quad M=4.28^{\pm 0.10}_{\pm 0.21}\,\times 10^6\,M_{\odot},\quad D=8.32^{\pm 0.07}_{\pm 0.14}\, kpc, \quad \tx{and}\quad \theta_{d}=48.7\,\pm\,7 \mu as.\label{data}
\ea                     
Here $M_{\odot}$ is the mass of the Sun. Having laid the theoretical foundation, we move to gauge the imprint of magnetic charge and DM halo on the lensing observables (\r{obs}).
\begin{figure}[H]
\begin{center}
\begin{tabular}{cc}
\includegraphics[width=0.4\columnwidth]{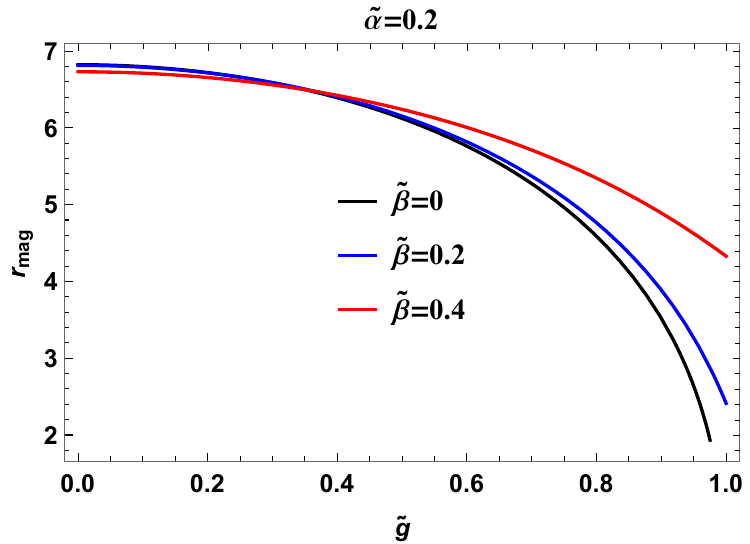}&
\includegraphics[width=0.4\columnwidth]{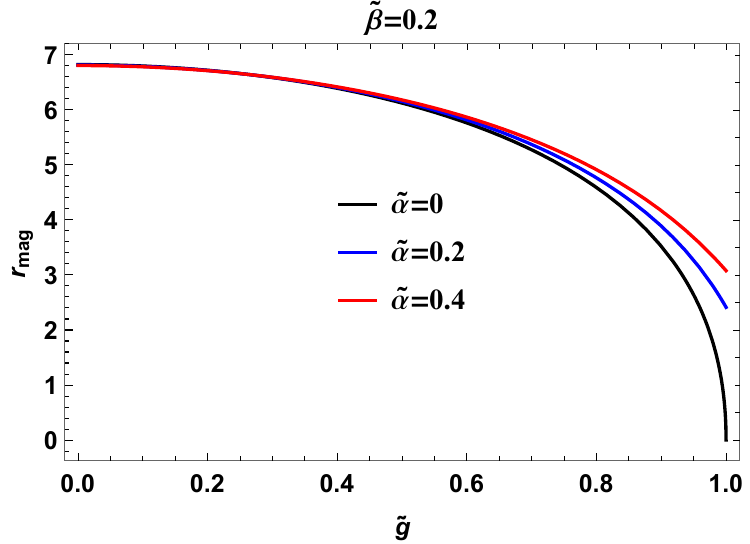}
\end{tabular}
\caption{Variation of relative magnification with $\tg$ for different values of $\tb$ (left panel) and different values of $\ta$ (right panel). }\label{rmag1}
\end{center}
\end{figure}
\begin{figure}[H]
\begin{center}
\begin{tabular}{cc}
\includegraphics[width=0.4\columnwidth]{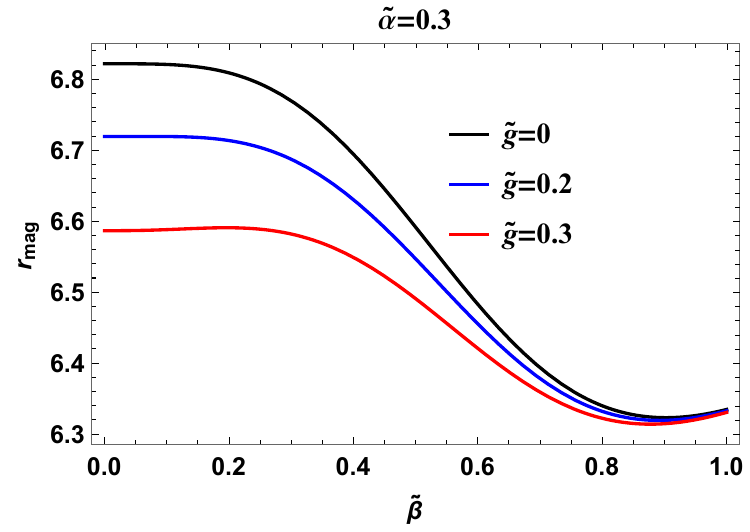}&
\includegraphics[width=0.4\columnwidth]{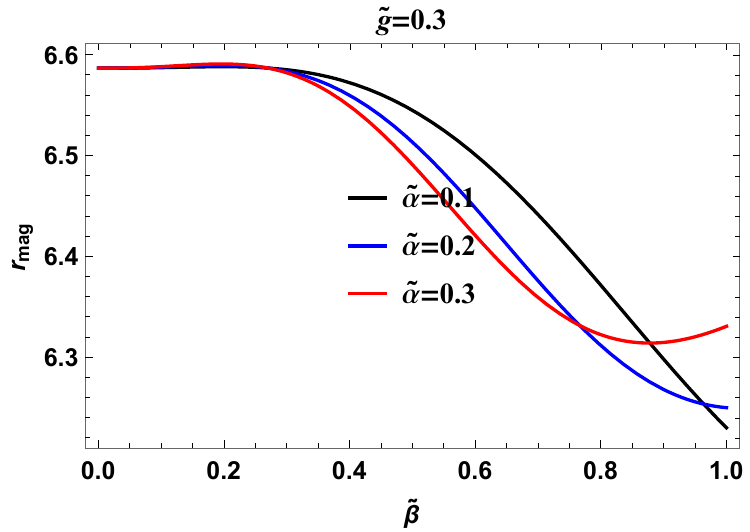}
\end{tabular}
\caption{Variation of relative magnification with $\tb$ for different values of $\tg$ (left panel) and different values of $\ta$ (right panel). }\label{rmag2}
\end{center}
\end{figure}
\begin{figure}[H]
\begin{center}
\begin{tabular}{cc}
\includegraphics[width=0.4\columnwidth]{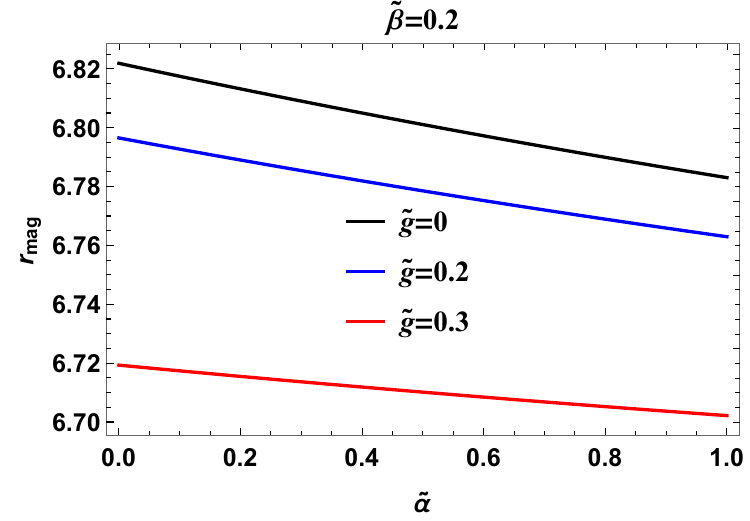}&
\includegraphics[width=0.4\columnwidth]{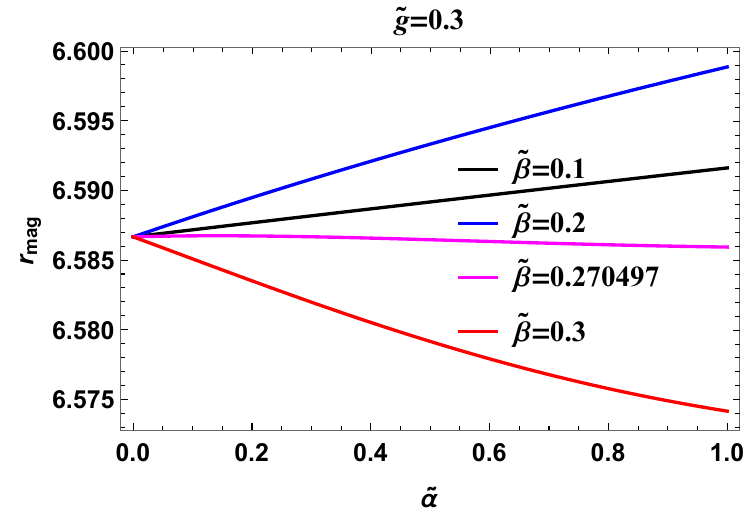}
\end{tabular}
\caption{Variation of relative magnification with $\ta$ for different values of $\tg$ (left panel) and different values of $\tb$ (right panel). }\label{rmag3}
\end{center}
\end{figure}   
Since $r_{mag} \propto \f{1}{\ab}$, the effect of charge and halo parameters on the relative magnification should be reverse as that of $\ab$ which is the case visible from Fig. (\r{rmag1}), (\r{rmag2}), and (\r{rmag3}). Fig. (\r{rmag3}) reveals, similar to the $\ab$ case, nullification of DM effect at $\tb_{ca}=0.270497$ when $\tg=0.3$. The relative magnification increases with $\ta$ when $\tb\,<\,\tb_{ca}$. Next, by modelling SMBH $SgrA^*$ as an MHDM BH, we elucidate the combined effect of NED and DM halo on the angular separation $s$ and angular position $\theta_{\infty}$.      
\begin{figure}[H]
\begin{center}
\begin{tabular}{cc}
\includegraphics[width=0.4\columnwidth]{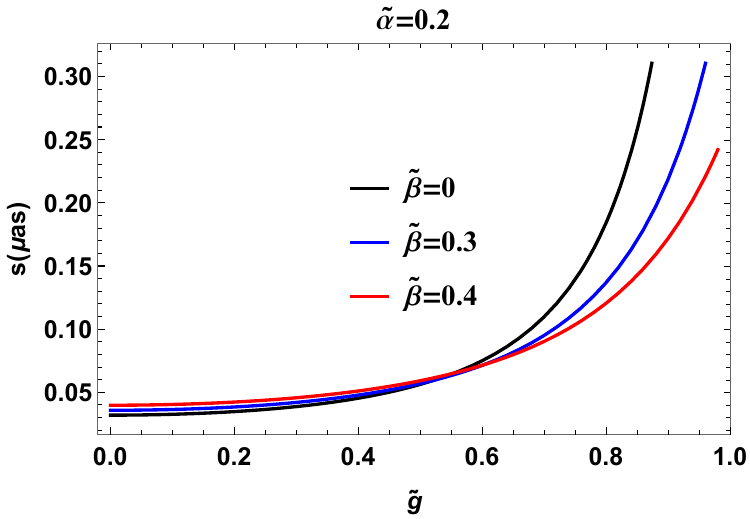}&
\includegraphics[width=0.4\columnwidth]{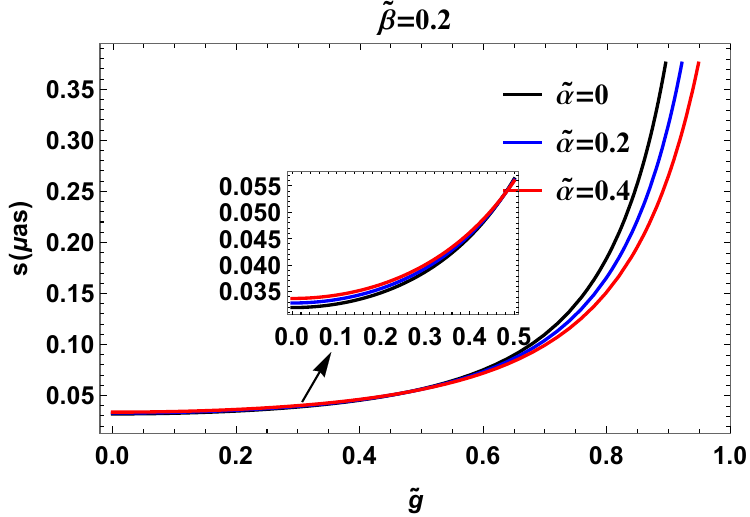}
\end{tabular}
\caption{Variation of angular separation with $\tg$ for different values of $\tb$ (left panel) and different values of $\ta$ (right panel). }\label{s1}
\end{center}
\end{figure}
\begin{figure}[H]
\begin{center}
\begin{tabular}{cc}
\includegraphics[width=0.4\columnwidth]{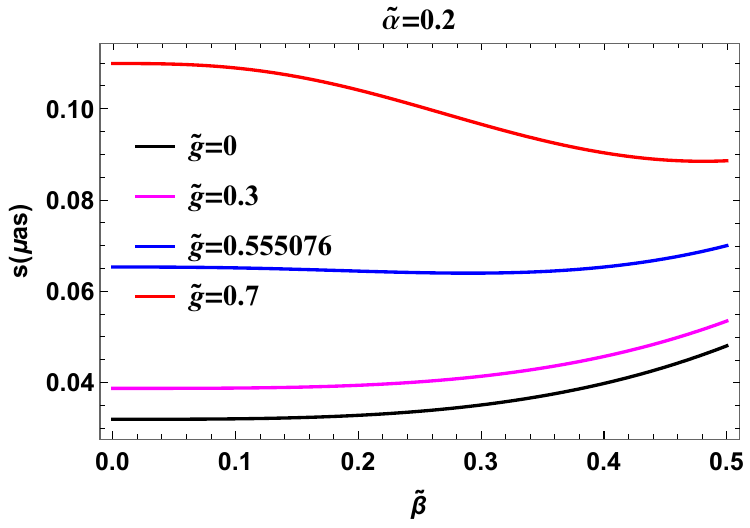}&
\includegraphics[width=0.4\columnwidth]{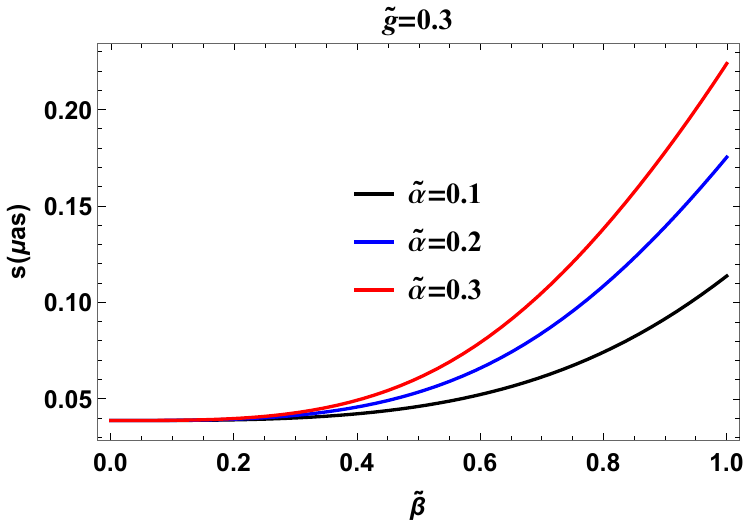}
\end{tabular}
\caption{Variation of angular separation with $\tb$ for different values of $\tg$ (left panel) and different values of $\ta$ (right panel). }\label{s2}
\end{center}
\end{figure}
\begin{figure}[H]
\begin{center}
\begin{tabular}{cc}
\includegraphics[width=0.4\columnwidth]{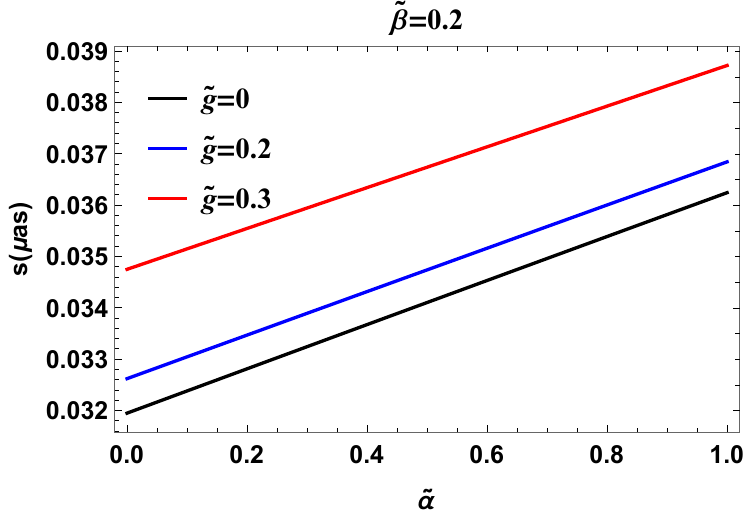}&
\includegraphics[width=0.4\columnwidth]{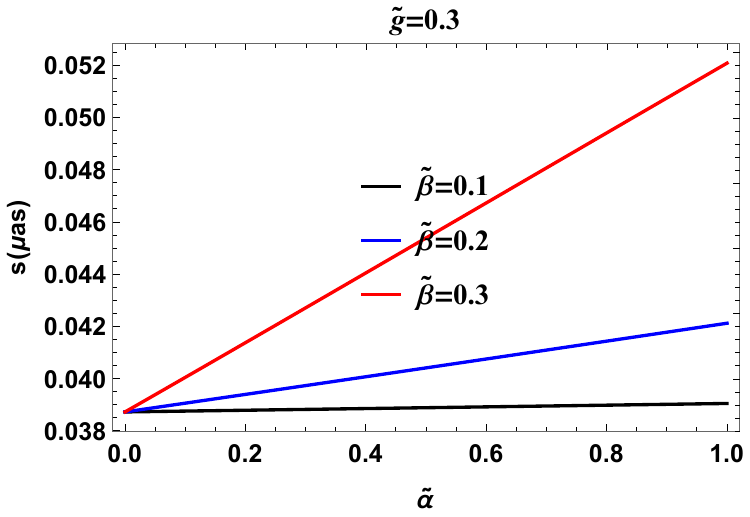}
\end{tabular}
\caption{Variation of angular separation with $\ta$ for different values of $\tg$ (left panel) and different values of $\tb$ (right panel). }\label{s3}
\end{center}
\end{figure}   
Fig. (\r{s1}), (\r{s2}), and (\r{s3}) shed light on the angular separation's dependence on the charge and halo parameters. The angular separation for an MHDM BH is always larger than that for a \s BH. Fig. (\r{s1}) reveals a critical charge $\tg_{cs}$ where the competing effect of halo parameters $\ta$ and $\tb$ cancel each other, leaving only the imprint of the magnetic charge on the angular separation. At $\ta=0.2$, we have $\tg_{cs}=0.555706$. Fig. (\r{s2}) demonstrates the fact that when $\tg\,<\,\tg_{cs}$, the angular separation increases with $\tb$, albeit slowly, and for $\tg\,>\,\tg_{cs}$, the angular separation decreases with $\tb$.
\begin{figure}[H]
\begin{center}
\begin{tabular}{cc}
\includegraphics[width=0.4\columnwidth]{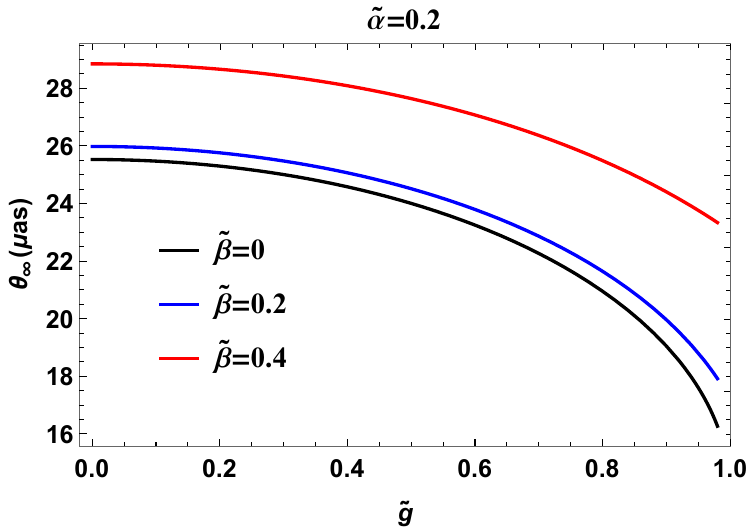}&
\includegraphics[width=0.4\columnwidth]{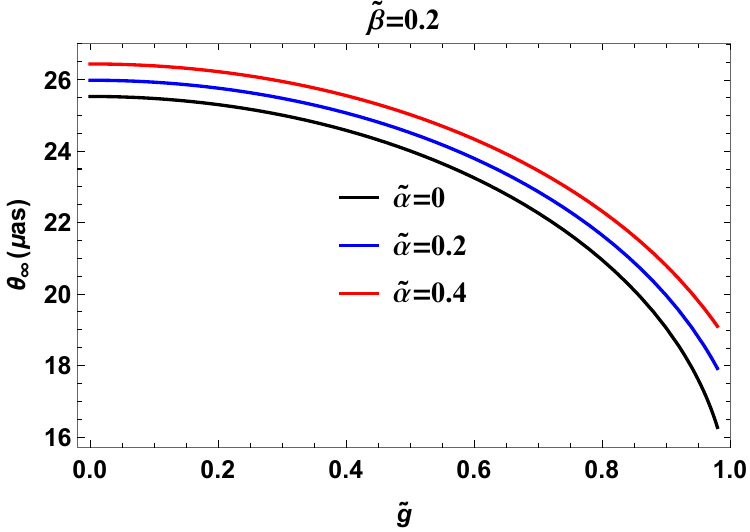}
\end{tabular}
\caption{Variation of angular position $\theta _\infty$ with $\tg$ for different values of $\tb$ (left panel) and different values of $\ta$ (right panel). }\label{tin1}
\end{center}
\end{figure}
\begin{figure}[H]
\begin{center}
\begin{tabular}{cc}
\includegraphics[width=0.4\columnwidth]{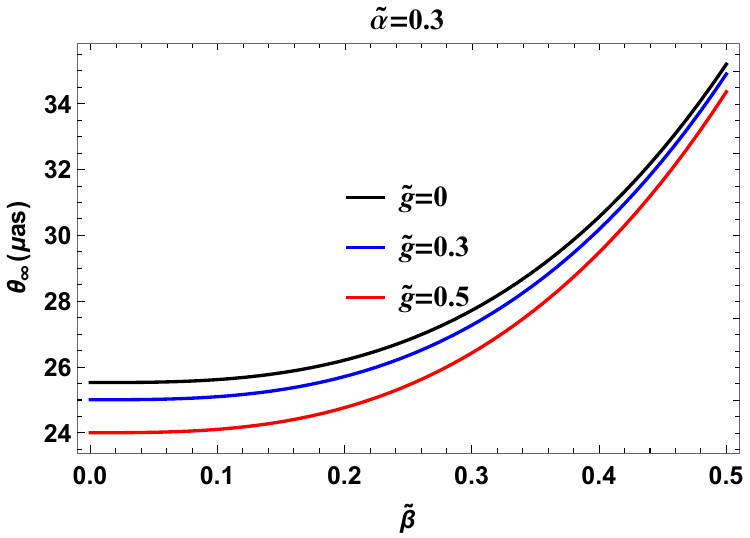}&
\includegraphics[width=0.4\columnwidth]{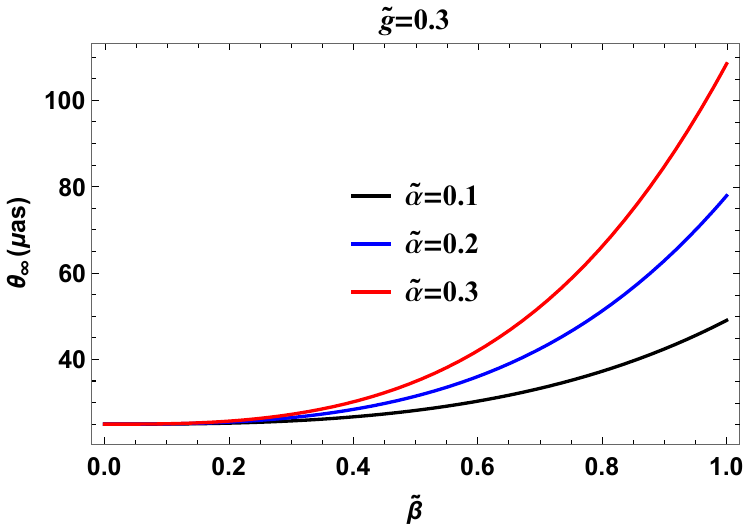}
\end{tabular}
\caption{Variation of angular position $\theta _\infty$ with $\tb$ for different values of $\tg$ (left panel) and different values of $\ta$ (right panel). }\label{tin2}
\end{center}
\end{figure}
\begin{figure}[H]
\begin{center}
\begin{tabular}{cc}
\includegraphics[width=0.4\columnwidth]{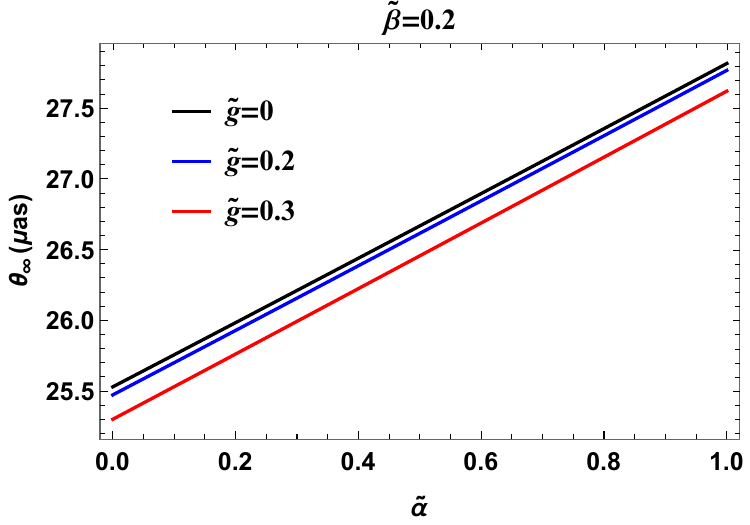}&
\includegraphics[width=0.4\columnwidth]{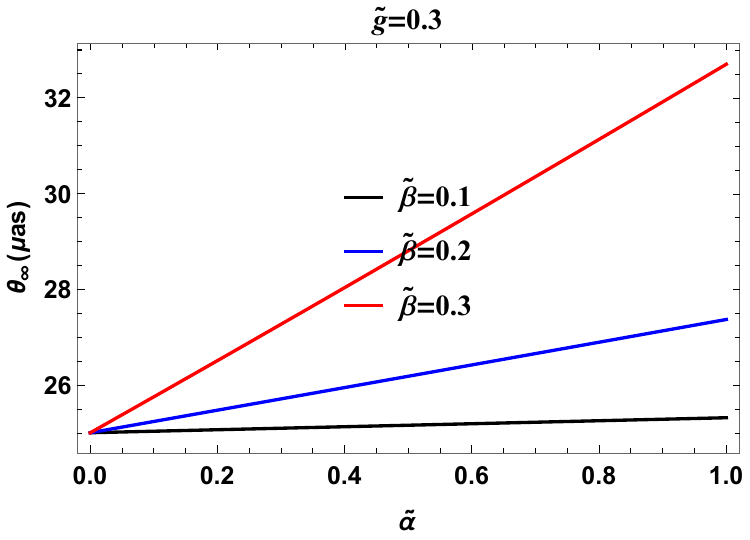}
\end{tabular}
\caption{Variation of angular position $\theta _\infty$ with $\ta$ for different values of $\tg$ (left panel) and different values of $\tb$ (right panel). }\label{tin3}
\end{center}
\end{figure}   
Fig. (\r{tin1}) reveals an adverse impact of magnetic charge on the angular position $\theta _\infty$, whereas Fig. (\r{tin2}) and (\r{tin3}) unravel a favourable impact of DM halo parameters on $\theta _\infty$. Having explored lensing observables of an MHDM BH, we now move to obtain parameter bounds that will make our BH model consistent with experimental observations.
\section{parameter estimation of an MHDM bh}
Theoretical predictions along with experimental observations help us extract bounds on the free parameters and enable us to refute or accept a proposed model. We will utilize the experimental results (\r{data}) and the method illustrated in \c{chi} to constrain parameters $g$, $\a$, and $\b$ that will make our model concordant with experimental observations within $1\sigma$ bounds. We use the following expression of $\chi^2$:
\be
\chi^2=\sum_{i} \f{\left\{\theta^{ex}_{di}-\theta^{th}_{di}(M,D,g,\a,\b)\right\}^{2}}{\sigma^{2}_{i}},\label{chi1}
\ee  
where $\theta^{ex}_{di}$ is the experimentally observed angular diameter of the $i^{th}$ BH, $\theta^{th}_{di}$ is its theoretically predicted value, and $\sigma_{i}$ is the error in the experimental observations. We, for our analysis, divide parameters $(M,D,g,\a,\b)$ into two cases - one is the interesting parameters $g,\,\a,\,\b$ and another is the uninteresting parameters $M,\,D$. We first minimize $\chi^2$ with respect to $M$ and $D$ over their range of values experimentally observed for a fixed value of $(g,\,\a,\,\b)$ and picking an SMBH. We repeat the process for the same $(g,\,\a,\,\b)$ but with a different SMBH. This process of minimization with respect to $M$ and $D$ is carried out for other combinations of $(g,\,\a,\,\b)$ such that $0\leq g \leq M$, $0\leq \a M^2 \leq 1 $, and $0 \leq \b \leq M$. We pick the combination $(g_m,\,\a_m,\,\b_m)$ that yields the minimum value of $\chi^2$ $(\chi_{m}^{2})$. Upper bound, within $1\sigma$ confidence level, on one interesting parameter is then obtained by solving the equation $\chi^2=\chi^{2}_{m}+3.50$ by keeping the other two parameters fixed at their values obtained for $\chi_{m}^{2}$. For our model, $\chi^2$ gets minimized at $(0.2446M,\,0.4254/ M^2,\,0.2182M)$. The upper bounds of $g,\,\a,\,\b$ are found to be as follows:
\be
g:\quad 0.7572M \quad \a M^2:\quad 1.46963 \quad \b: \quad 0.336436M.
\ee
Our analysis thus makes it amply clear that the model proposed in this manuscript can serve as a viable candidate for an SMBH.
\section{conclusions}
With a view to exploring the interplay of NED and DM halo, we first obtained a static and spherically symmetric metric of a BH in the \h where the BH's magnetic charge arises from NED. To this end, the metric for a pure \h reported earlier was utilized. Extracting expressions for the energy density and pressures for the pure \h and employing the Lagrangian density for NED, we solved modified field equations, which ultimately led us to the desired metric. It is interesting to note that even though the metric components do not diverge at $r=0$, the metric is not regular there, which is made clear by the divergence of scalar invariants such as the Kretschmann Scalar at $r=0$. This makes the singularity at $r=0$ an irremovable one. Our analysis revealed the adverse effect of magnetic charge on the event horizon, whereas the halo parameters were found to impact it favourably. At some combinations of $(g,\,\a,\,\b)$, competing effects of charge and halo parameters cancel each other, leaving the event horizon unaffected and at the position as that of a \s BH. Some such combinations are $(0.154M,\, 0.15/M^2,\,0.15M)$, $(0.2M,\,0.2/M^2,\,0.1625M)$, and $(0.1M,\,0.21/M^2,\,0.1M)$. At these values, it will be impossible to distinguish between a \s BH and an MHDM BH with respect to the positions of their event horizons.\\
We then moved on to study strong GL with the intent to gauge how the combined system of NED and DM halo effects lensing observables. The influence of charge and halo parameters on the critical impact parameter is similar to the horizon case, and like the horizon case, we have found combinations $(\tg,\,\ta,\,\tb)$ where the charge and halo effects cancel each other. Some such combinations are $(0.3495,\,0.3,\,0.2)$, $(0.2,\,0.2283,\, 0.15)$, and $(0.3,\,0.2,\, 0.2075)$. We elucidated the impact of charge and halo parameters on the lensing coefficients $\ab$, $\bb$, and on the deflection angle $\gamma_{D}$ graphically. In case of the lensing coefficient $\ab$, there exists, for a fixed value of $\tg$, a critical value of $\tb$ $(\tb_{ca})$ where the halo effect is absent and only the imprint of charge remains over $\ab$. One such value is $\tb_{ca}=0.270497$ at $\tg=0.3$ and $\tb_{ca}=0.00111479$ at $\tg=0$. In such a scenario a (un)charged BH has the same $\ab$ irrespective of whether it is in a halo or not. Likewise, we find MHDM BHs with a particular set of $(\tg,\,\ta,\,\tb)$ that deflect a light ray by the same angle as that of a \s BH as a result of the diminishing effect produced by the charge being cancelled by the augmenting effect of halo parameters. Some such cases are $(0.2,\,0.2,\,0.157153)$, $(0.1,\,0.18842,\,0.1)$, and $(0.0727999,\,0.1,\,0.1)$.\\
Our exposition into lensing observables such as angular separation $s$, relative magnification $r_{mag}$, and angular position $\theta_{\infty}$ revealed a significant impact of charge and DM parameters on them. Since $r_{mag} \propto \f{1}{\ab}$, the influence of our model parameters on the relative magnification is opposite to that of the $\ab$ case, which is clearly illustrated in Fig. (\r{rmag1}), (\r{rmag2}), and (\r{rmag3}). The angular separation for an MHDM BH is always found to be greater than that of a Schwarzschild BH, thereby making it possible to resolve the outer bright ring and the inner closely packed ring easily. The angular position $\theta_{\infty}$ displayed similar dependence on model parameters as that of critical impact parameter since $\theta_{\infty} \propto b_m$. Finally, employing the prescription illustrated in \c{chi} and utilizing experimental observations tabulated in Table (\r{data}), we extracted bounds on the model parameters within $1\sigma$ confidence level, which came out to 0.7572M for $g$, $1.46963/M^2$ for $\a$, and $0.336436M$ for $\b$. These results make an MHDM BH a viable candidate for an SMBH. Observations related to quasiperiodic oscillations of quasars can also be utilized to test a proposed model. This is our topic of interest for future endeavour. 

\end{document}